\documentclass{webofc}
\usepackage[varg]{txfonts}   
\usepackage{lipsum}
\usepackage{xcolor}
\usepackage{indentfirst}
\begin{document}
\title{The advantage of Bolometric Interferometry for controlling Galactic foreground contamination in CMB primordial \textit{B}-modes measurements}
%
%

\author{\lastname{E.~Manzan}\inst{1,2}\fnsep\thanks{\email{elenia.manzan@unimi.it}} \and
        \lastname{M.~Regnier}\inst{3} \and
        \lastname{J-Ch.~Hamilton}\inst{3} \and
	 \lastname{A.~Mennella}\inst{1,2} \and
	\lastname{J.~Errard}\inst{3} \and 
	\lastname{L.~Zapelli}\inst{1,2} \and 
	\lastname{S.A.~Torchinsky}\inst{3,4} \and
	\lastname{S.~Paradiso}\inst{5,6} \and
	\lastname{E.~Battistelli} \inst{7} \and
	\lastname{M.~Bersanelli}\inst{1,2} \and
       \lastname{P.~De Bernardis}\inst{7} \and
       \lastname{M.~De Petris} \inst{7} \and
       \lastname{G.~D’Alessandro} \inst{7} \and
       \lastname{M.~Gervasi} \inst{8} \and
       \lastname{S.~Masi} \inst{7} \and
       \lastname{M.~Piat} \inst{3} \and
       \lastname{E.~Rasztocky} \inst{9} \and
       \lastname{G.E~Romero} \inst{9} \and
       \lastname{C.G.~Scoccola} \inst{10, 11} \and
       \lastname{M.~Zannoni} \inst{8} \and
	the QUBIC Collaboration
}

\institute{Università degli studi di Milano, Italy
\and
	INFN sezione di Milano, Milano, Italy
\and
	Université Paris Cité, CNRS, Astroparticule et Cosmologie, Paris, France
\and
	Université PSL, Observatoire de Paris, AstroParticule et Cosmologie, Paris, France
\and
	Waterloo Centre for Astrophysics, University of Waterloo, Waterloo, Canada
\and
	Department of Physics and Astronomy, University of Waterloo, Waterloo, Canada
\and
	Università di Roma - La Sapienza, Italy
\and
	Università di Milano – Bicocca, Italy
\and
	Instituto Argentino de Radioastronomía, Buenos Aires, Argentina
\and
	Consejo Nacional de Investigaciones Científicas y Técnicas, Ciudad de Buenos Aires, Argentina
\and
	Facultad de Ciencias Astronómicas y Geofísicas, Universidad Nacional de La Plata, Argentina
}

\abstract{
In the quest for the faint primordial \textit{B}-mode polarization of the Cosmic Microwave Background, three are the key requirements for any present or future experiment: an utmost sensitivity, excellent control over instrumental systematic effects and over Galactic foreground contamination.

Bolometric Interferometry (BI) is a novel technique that matches them all by combining the sensitivity of bolometric detectors, the control of instrumental systematics from interferometry and a software-based, tunable, in-band spectral resolution due to its ability to perform band-splitting during data analysis (spectral imaging). 

In this paper, we investigate how the spectral imaging capability of BI can help in detecting residual contamination in case an over-simplified model of foreground emission is assumed in the analysis. To mimic this situation, we focus on the next generation of ground-based CMB experiment, CMB-S4, and compare its anticipated sensitivities, frequency and sky coverage with a hypothetical version of the same experiment based on BI, CMB-S4/BI, assuming that line-of-sight (LOS) frequency decorrelation is present in dust emission but is not accounted for during component separation.

We show results from a Monte-Carlo analysis based on a parametric component separation method (FGBuster), highlighting how BI has the potential to diagnose the presence of foreground residuals in estimates of the tensor-to-scalar ratio $r$ in the case of unaccounted Galactic dust LOS frequency decorrelation.
}

\maketitle
%

\section{Introduction}
\label{sec-intro}

As the current best upper limit on the tensor-to-scalar ratio $r$ of the CMB primordial \textit{B}-modes tightens to $r<0.032$ \cite{Tristram_2022}, measuring such a faint signal requires instruments with utmost sensitivity, excellent control over instrumental systematic effects and over foreground contamination. The latter is achieved through multifrequency observations, to exploit the different spectral behaviour of the blackbody CMB emission with respect to the Galactic synchrotron and dust grains emissions, which are conventionally assumed to scale over frequency as a power-law and a modified blackbody (MBB), respectively. However, since both synchrotron and thermal dust are brighter than the CMB emission in polarization, modern CMB experiments are also relying, now more than ever, on improved foreground models to mitigate as much as possible the residuals due to improper or overly simplified modeling.

Indeed, over the years many models have been implemented in PySM \cite{Thorne_2017} (Python Sky Model) to account for additional complexities in the Galactic dust emission, such as different dust grain compositions \cite{Hensley_2017}, the superimposition of coherently aligned molecular clouds with different temperatures and spectral indices \cite{Finkbeiner_1999,Martinez_Solaeche_2018} or line-of-sight (LOS) frequency decorrelations caused by a frequency-dependent polarization angle due to misalignments of various molecular clouds along the LOS, following the statistical approach described in \cite{Vansyngel_2018}.

In this paper, we investigate how the spectral imaging capability of an unconventional technique for CMB polarimetry, Bolometric Interferometry (BI) \cite{2020.QUBIC.PAPER1}, can help in diagnosing the presence of foreground residuals in case of improper or overly simplified foreground modeling. Precisely, we assume that LOS frequency decorrelations are present in the dust emission but are not accounted for during component separation and we compare the estimated tensor-to-scalar ratio, $r$, after a Monte-Carlo analysis based on a parametric component separation, in the case of the next generation of ground-based CMB experiment, CMB-S4 \cite{Abazajian_2022}, and of a hypothetical version of the same experiment based on BI, CMB-S4/BI.

\section{Bolometric Interferometry in a nutshell}
\label{sec-BI}

\begin{figure}[h]
\centering
\includegraphics[scale=0.16]{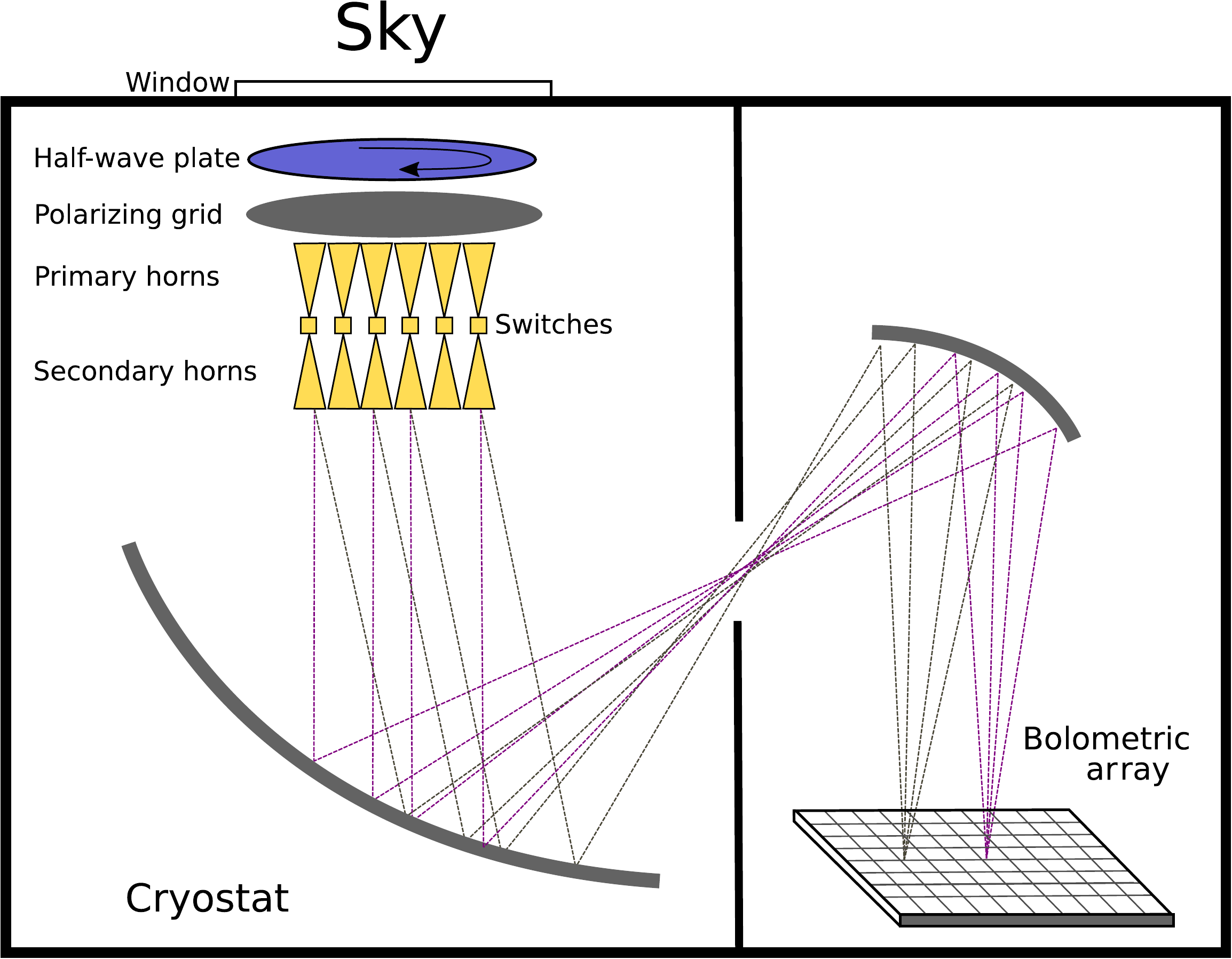} \quad
\includegraphics[scale=0.37]{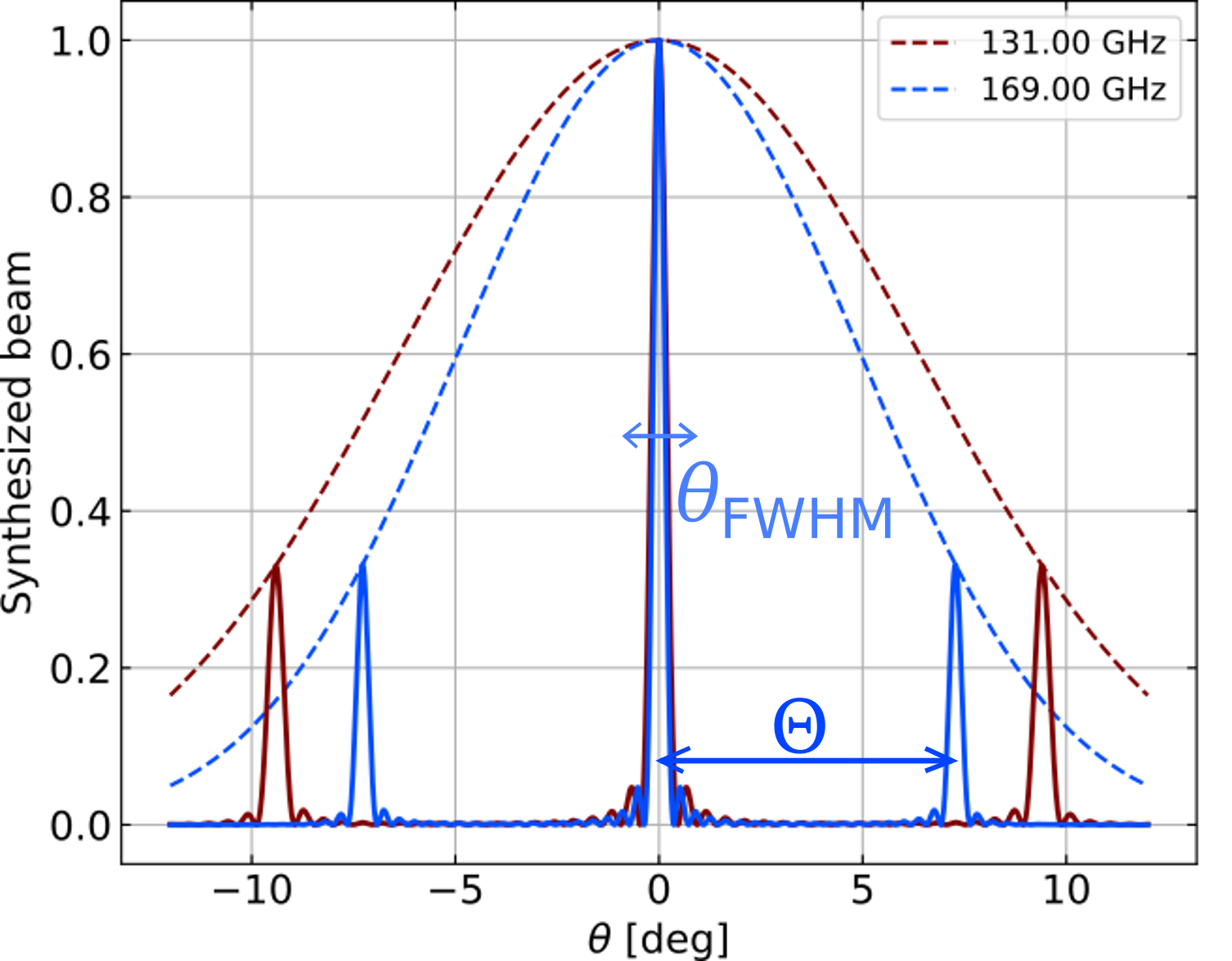}
\caption{\textit{Left panel}: schematic of a BI instrument. The sky signal is received by an array of back-to-back antennas and re-imaged onto the bolometric focal planes where the field interferes additively. A polarizer and a rotating half-wave plate make the instrument sensitive to linear polarization. \textit{Right panel}: azimuth cut of the {\em synthesized beam} (solid lines) at 131 GHz (red line) and at 169 GHz (blue line) for a detector at the center of the focal plane. Dashed lines represent the beam pattern of a single feed-horn antenna. One can appreciate the frequency-dependent position of the secondary peaks.}
\label{fig-BI_scheme}       
\end{figure}

Bolometric Interferometry combines the use of bolometers, which are state-of-the-art cryogenic broadband detectors providing high sensitivity, with the control of instrumental systematic effects typical of interferometry \cite{1981MNRAS.196.1067C,Bigot-Sazy2013}. A schematic of BI is shown in the left panel of Fig.~\ref{fig-BI_scheme}: an array of antennas receives the sky signal and re-emits it through a back array of twin antennas towards an optical combiner that focuses the radiation onto the focal plane, conversely to a traditional imager where each antenna is coupled to a single detector instead. Therefore, the signal received by each detector is the sky signal convolved with the joint angular response of all the antennas simultaneously, which we call \textit{synthesized beam}, and exhibits a multiple-peaked shape as shown in the right panel of Fig.~\ref{fig-BI_scheme}.

The interested reader can find more details on BI and on its current state-of-the-art, represented by the QUBIC experiment, in \cite{2020.QUBIC.PAPER1, 2020.QUBIC.PAPER2,2020.QUBIC.PAPER3,2020.QUBIC.PAPER4, 2020.QUBIC.PAPER5, 2020.QUBIC.PAPER6, 2020.QUBIC.PAPER7, 2020.QUBIC.PAPER8}.

\subsection{Spectral Imaging}
\label{sec-Spectral_Imaging}

The width of the peaks in the synthesized beam, $\theta_\mathrm{FWHM}$, and their angular distance, $\Theta$, depend on the signal wavelength, $\lambda$, on the number of antennas along the maximum axis of the antenna array, $P$, and on the separation between two consecutive antennas, $\Delta h$, as \cite{2020.QUBIC.PAPER2}:

\begin{equation}
\label{eq_synth_beam_properties_dependece}
	\theta_\mathrm{FWHM}  = \frac{\lambda}{(P-1)\Delta h}, \,\,\,\,\,\,\,\,\,\,
	\Theta = \frac{\lambda}{\Delta h}.
\end{equation}

The linear dependence of the angular separation $\Theta$ on the wavelength $\lambda$ causes the position of the secondary peaks to significantly move as a function of the frequency, as shown in the right panel of Fig.~\ref{fig-BI_scheme}. Therefore, during data analysis one can recover the sky signal for two frequencies$, \nu_1$ and $\nu_2$, within the instrument frequency band as long as the secondary peaks are well-resolved. This occurs if $\Theta(\nu_2) -  \Theta(\nu_1) > \theta_\mathrm{FWHM}(\sqrt{\nu_1\nu_2})$, namely if $\frac{\Delta \nu}{\nu} \geq \frac{1}{P-1}$.

We call this technique \textit{spectral imaging} and, since it happens at the data analysis level, it allows us to tune the spectral resolution and also to re-analize the data with various spectral configurations to search for biases.

\section{Simulation set-up}
\label{sec-Simulation}

\subsection{Sky models}
\label{sec-Sky models}

In our simulation we consider three cases of dust emissions: i. a MBB with spatially varying temperature, $T_d(\hat{n})$, and spectral index, $\beta_d(\hat{n})$, modelled using the PySM \textbf{d1} model as

\begin{equation}
\label{eq:MBB}
I_d (\hat{n}, \nu)\ \bigg\rvert_{MBB} = A_{d,\nu_0}(\hat{n})\ \frac{B_{\nu}\left(T_d(\hat{n})\right)}{B_{\nu_0}\left(T_d(\hat{n})\right)}\left( \frac{\nu}{\nu_0} \right)^{\beta_d(\hat{n})},
\end{equation}
ii. a MBB with constant temperature, $T_d$, and spectral index, $\beta_d$, derived from Equation~\ref{eq:MBB} using the \textbf{d0} model and iii. a deviation from a single MBB that accounts for LOS frequency decorrelation, computed using the \textbf{d6} model \cite{Vansyngel_2018} by multiplying the \textbf{d1} emission in Equation~\ref{eq:MBB} for a decorrelation factor, $D$, as
\begin{equation}
I_d (\hat{n}, \nu) = D(\nu, \nu_0, \ell_{corr})\ I_d (\hat{n}, \nu)\ \bigg\rvert_{MBB}
\end{equation}

\begin{figure}[h]
\centering
\sidecaption
\includegraphics[width=7cm, clip]{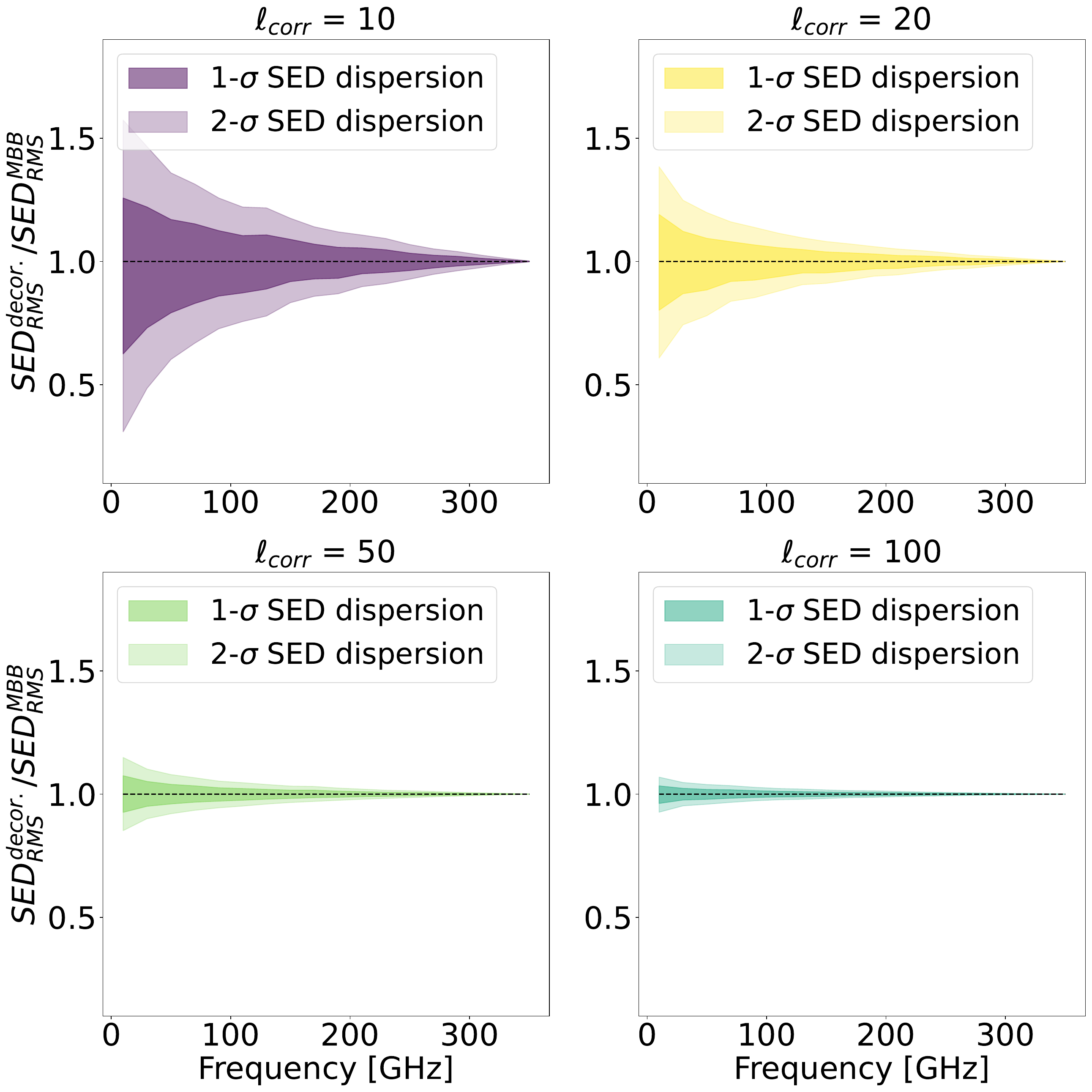}
\caption{Dispersion of the dust SED for different correlation lengths of the PySM \textbf{d6} model normalized by the single MBB emission (\textbf{d1} model). The colored areas represent the statistical deviation from a MBB for a given correlation length, evaluated over 500 realizations. As one can see, smaller correlation lengths $\ell_{corr}$ cause larger decorrelation. Moreover, for a given $\ell_{corr}$, the decorrelation decreases as $\nu$ gets closer to $\nu_0$, where $\nu_0 = 353$ GHz for polarization, as described in \cite{Vansyngel_2018}.}
\label{fig-d6_dispersion}       
\end{figure}

The decorrelation factor, $D(\nu, \nu_0, \ell_{corr})$, is a function of the simulated frequency $\nu$, a reference frequency $\nu_0$ for which $D(\nu = \nu_0) = 1$, and of the correlation length $\ell_{corr}$, a parameter that quantifies the deviation from a single MBB. For each simulated frequency $\nu$, the decorrelation factor $D$ is randomly sampled from a Gaussian distribution with center $\mu = 1 $ and standard deviation $\sigma$ that scales as the inverse of the correlation length, $\sigma=\sigma(1/\ell_{corr})$. Therefore, a smaller correlation length causes larger decorrelation, and viceversa, as it is shown in Fig.~\ref{fig-d6_dispersion}. This model mimics the presence of a frequency-varying polarization angle caused by some misalignments of the magnetic field (and therefore of the molecular clouds) along the LOS, without making any physical assumption on the misalignment itself.

For the synchrotron, we use the \textbf{s0} and \textbf{s1} models, which simulate a power-law with spatially constant or varying spectral index, $\beta_s$, respectively.

\subsection{Instrumental configurations}
\label{sec-Instruments}

\begin{figure}[h]
\includegraphics[scale=0.17]{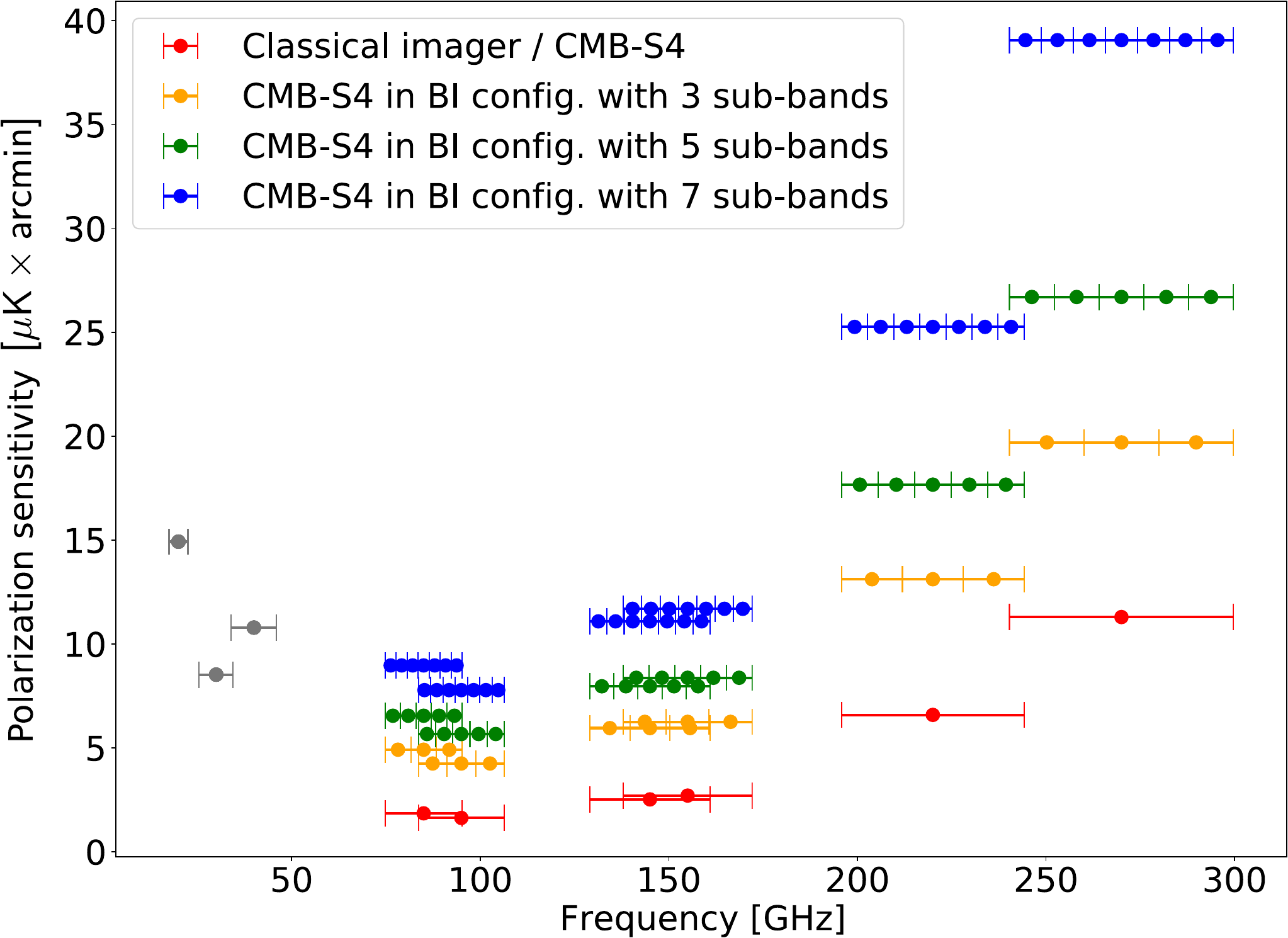} \quad
\includegraphics[scale=0.3]{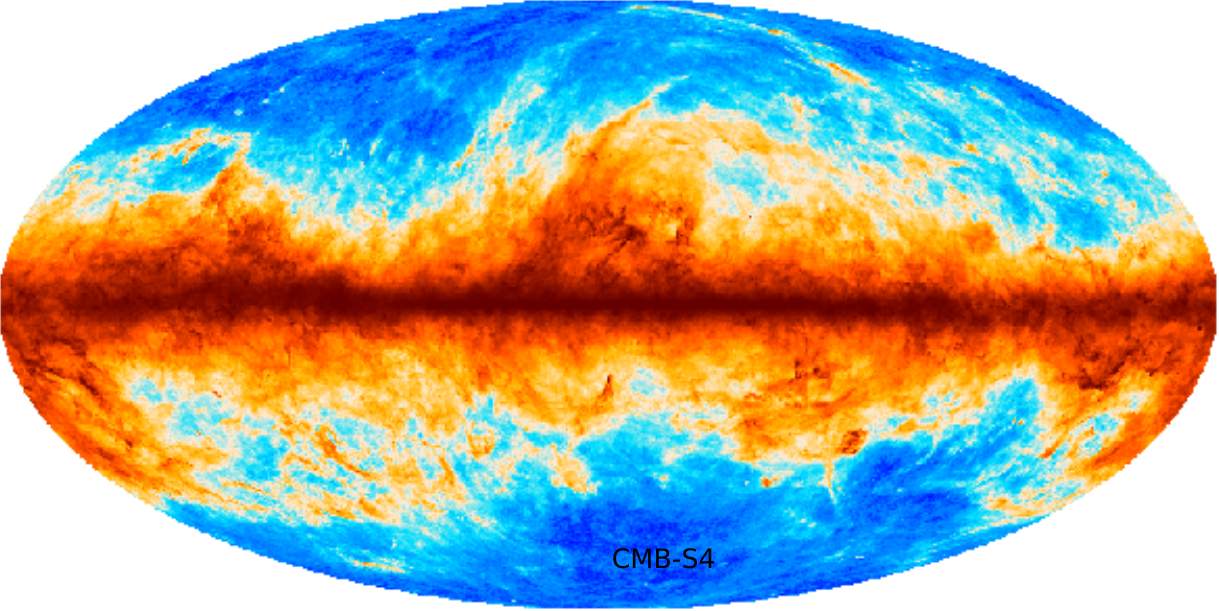}
\caption{\textit{Left Panel}: Polarization sensitivity of CMB-S4 and three examples of CMB-S4/BI, with $n_\mathrm{sub}=3,5,7$ respectively. The three lowest frequency bands in grey are identical for all the instruments because we choose not to split the synchrotron dominated frequency channels. \textit{Right Panel}: The anticipated CMB-S4 sky patch (white patch), compared to the dust emission map at 150 GHz plotted underneath.}
\label{fig-instrum_summary}       
\end{figure}

To simulate the next generation of ground-based CMB experiment, CMB-S4, we consider its anticipated nine frequency channels, $\Delta \nu_i$, sensitivities, $\sigma_{i}$, and 3$\%$ circular sky patch following \cite{Abazajian_2022}. To simulate a BI version of CMB-S4, we assume to perform spectral imaging during data analysis on the six highest frequency channels where the dust emission dominates over the synchrotron. Precisely, we split each frequency band $\Delta \nu_i$ into a number of sub-bands $n_\mathrm{sub}$ and correspondingly increase the white noise level, $\sigma^\mathrm{BI}_{i}$, as
\begin{equation}
	\Delta \nu^\mathrm{BI}_i = \frac{\Delta \nu_i}{n_\mathrm{sub}} , \,\,\,\,\,\,\,\,\,\,
	\sigma^\mathrm{BI}_{i} = \sigma_i \times \sqrt{n_\mathrm{sub}} \times \varepsilon \quad \mathrm{for }\ i = 4, ..., 9.
	\label{bandwidth_and_sigma_BI}
\end{equation}
The \textit{sub-optimality} factor, $\varepsilon \in [1.2, 1.6]$, acts as a proxy that increases the white noise level to account for the BI correlated noise, as described in \cite{2020.QUBIC.PAPER2}, and $n_\mathrm{sub}$ ranges from 2 to 8, for a total of seven simulated CMB-S4/BI cases. The comparison between the CMB-S4 sensitivity and three cases of CMB-S4/BI is shown in Fig.~\ref{fig-instrum_summary}, along with the simulated sky patch.

\subsection{Simulation pipeline}
\label{sec-Pipeline}
Our simulation pipeline consists of 500 iterations of a Monte-Carlo chain. For each iteration, we generate a CMB realization, a noise realization for each frequency channel of each instrumental configuration and the foreground maps in one of the three cases: \textbf{d0s0}, \textbf{d1s1} or \textbf{d6s1} model. We apply a parametric component separation method to the input maps, assuming the dust model to be a MBB independently of the true input dust emission. In the \textbf{d0s0} and \textbf{d1s1} case, this corresponds to fitting with the correct model, whereas in the \textbf{d6s1} it mimics the situation in which our model is overly simplified due to our lack of knowledge of the details of the dust emission. We compute the cross-power spectra from each of the 500 reconstructed CMB maps and the corresponding (Gaussian) likelihood on $r$, from which we extract the maximum-likelihood value of $r$. 

In this paper we show the results obtained using the FGBuster component separation code \cite{Stompor}. However, we repeated the study with a different parametric code, Commander \cite{eriksen_2006}, and found coherent results. The interested reader can find more details in \cite{regnier2023identifying}.

\section{Results}
\label{sec-Results}

\begin{figure}[h]
\includegraphics[scale=0.28]{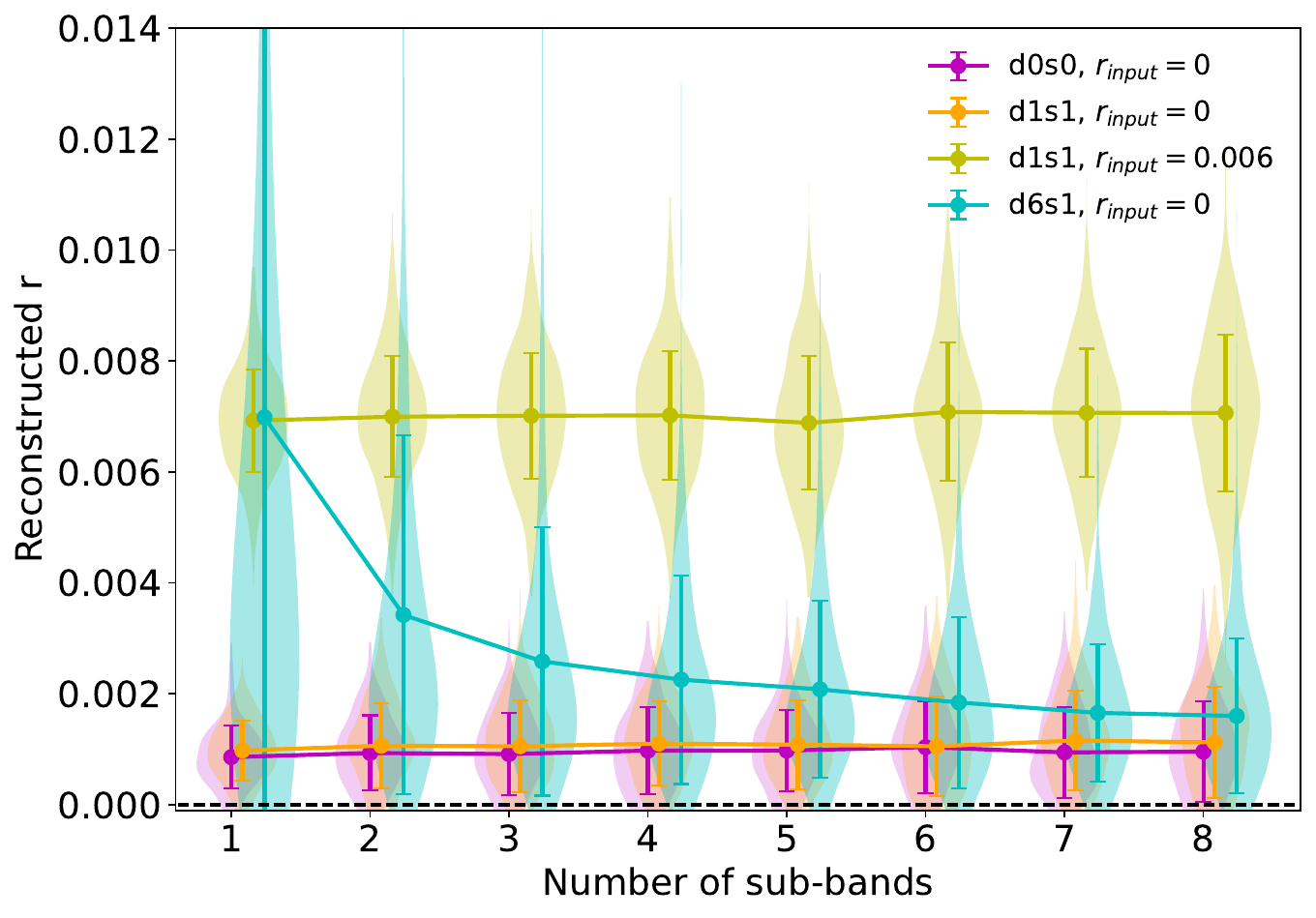} \quad
\includegraphics[scale=0.13]{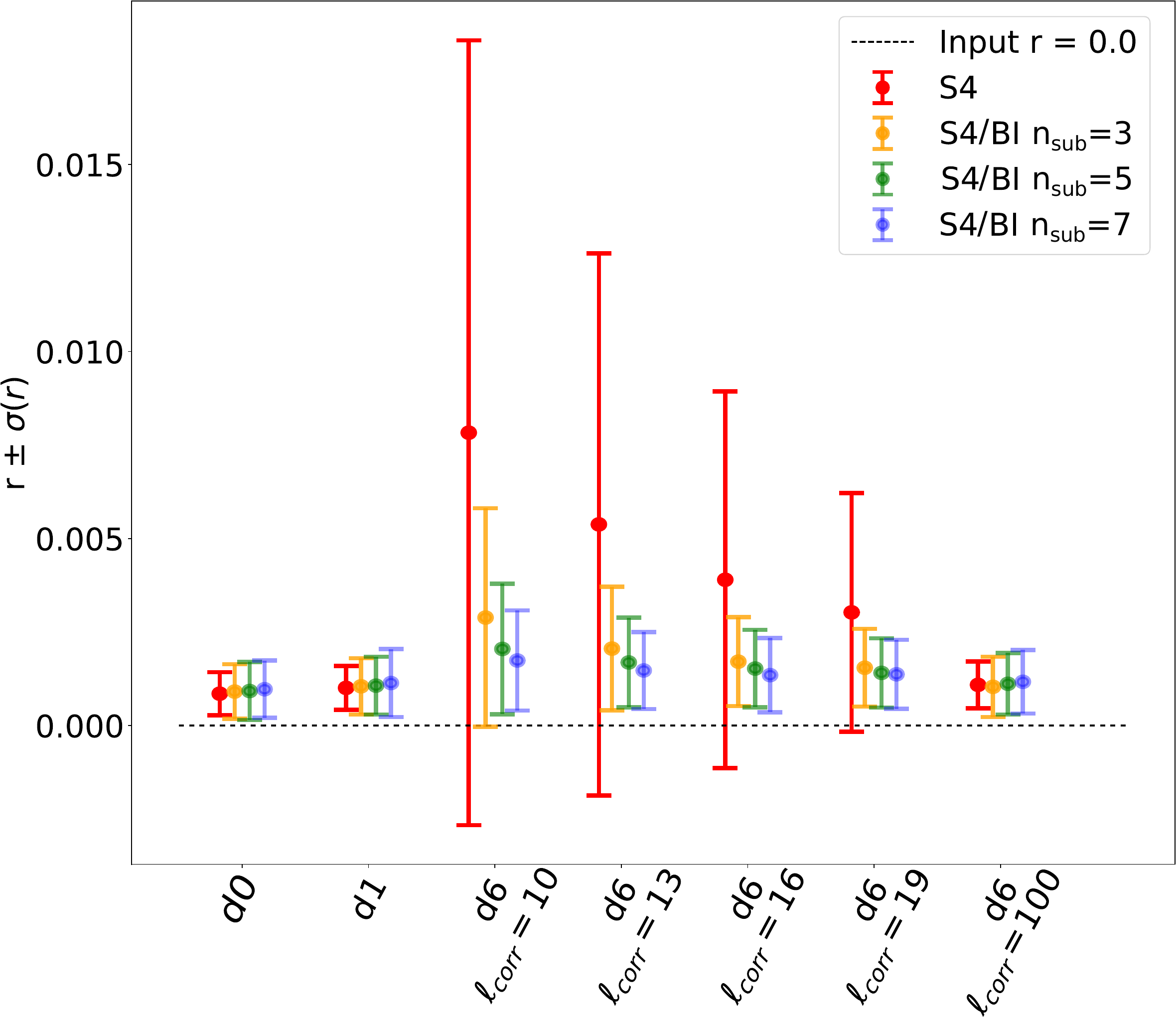} 
\caption{\textit{Left Panel}: Average maximum-likelihood value of $r$ and standard deviation as a function of the number of sub-bands, computed from the distribution shown vertically as a \textit{violin plot}. The cases where we fit for the correct model are shown in purple, orange and green (\textbf{d0} or \textbf{d1}, with input $r=0$ or $r=0.006$). The cyan curve shows the result when LOS frequency decorrelations are present in the dust emission (model \textbf{d6s1} with $\ell_\mathrm{corr}=10$) but not accounted for in the analysis. \textit{Right Panel}: Summary of the result on $r$ for all the simulated foreground models (\textbf{d0s0}, \textbf{d1s1} and several $\ell_\mathrm{corr}$ cases of \textbf{d6s1}) with input $r=0$. For ease of reading, only CMB-S4 and CMB-S4/BI with $n_\mathrm{sub}=3,5,7$ are shown.}
\label{fig-results}       
\end{figure}

The average maximum likelihood value of $r$ and standard deviation as a function of the number of sub-bands is shown in Fig.~\ref{fig-results}. A classical imager, like CMB-S4, would only recover the case $n_\mathrm{sub}=1$. In the left panel, our results show that when we fit for the correct model (\textbf{d0} or \textbf{d1}), we expect the result from BI to be independent of $n_\mathrm{sub}$\footnote{The small bias is due to an $E\rightarrow B$ modes leakage caused by the power spectra computation on a sky patch, where the spherical harmonics are no longer orthogonal. This bias could be mitigated by increasing the apodization radius of the mask at the expense of a smaller effective sky fraction ($<3\%$), but this optimization is outside the scope of the study.}. Instead, when dust LOS frequency decorrelations are present but not accounted for during component separation, a bolometric interferometer would recover a decreasing value of $r$ as a function of $n_\mathrm{sub}$: increasing the number of sub-bands therefore allows us to reduce the bias with respect to a traditional imager. Moreover, the decreasing trend as a function of $n_\mathrm{sub}$, conversely to the expected constant trend, is itself a direct hint of the contamination from foreground residuals. This advantage of BI over a classical imager is mantained also for smaller levels of decorrelation, as shown in the right panel of Fig.~\ref{fig-results}.

\section{Conclusions}
\label{sec-Conclusions}
In this paper we have shown that the spectral imaging capability of Bolometric Interferometry can help diagnosing the presence of foreground residuals in estimates of the tensor-to-scalar ratio $r$ when LOS frequency decorrelations are present in the dust emission but not properly accounted for during component separation, whereas an imager of similar performance could be biased. 

Although neglecting the impact of instrumental systematic effects, this study illustrates the potential of BI in the context of future CMB polarization experiments that will be challenged by complex foregrounds in their quest for \textit{B}-modes detection.

\bibliography{biblio}

\end{document}